\documentclass[%
 reprint,
 amsmath,amssymb,
 aps,
]{revtex4-1}

\usepackage{graphicx}
\usepackage{dcolumn}
\usepackage{bm}

\usepackage{balance}
\usepackage{mhchem}
\usepackage[LGR,T1]{fontenc}
\usepackage[latin9]{inputenc}
\usepackage{textcomp}
\usepackage{tipa}
\usepackage{amstext}
\usepackage{subscript}
\usepackage{times,mathptmx}
\usepackage{lastpage}
\usepackage{float}
\usepackage{fancyhdr}
\usepackage{fnpos}
\usepackage[english]{babel}
\usepackage{array}
\usepackage{droidsans}
\usepackage{charter}
\usepackage[T1]{fontenc}
\usepackage[usenames,dvipsnames]{xcolor}
\usepackage{setspace}
\usepackage[compact]{titlesec}
\usepackage{natbib}

\begin{document}


\title{Depth dependent decoherence caused by surface and external spins for NV centers in diamond}

\author{Wenlong Zhang, Jian Zhang, Junfeng Wang, Fupan Feng, Shengran Lin, Liren Lou, Wei Zhu\textit{$^{*}$} and Guanzhong Wang}

 \email{Corresponding authors: gzwang@ustc.edu.cn; zhuw@ustc.edu.cn}
\affiliation{%
 Key Laboratory of Strongly-Coupled Matter Physics, Chinese Academy of Sciences, and Hefei National Laboratory for Physical Science at Microscale, and Department of Physics, University of Science and Technology of China. Hefei, Anhui, 230026, P.R.China.\\
}%


\begin{abstract}
By efficient nanoscale plasma etching, the nitrogen-vacancy (NV) centers in diamond were brought to the sample surface step by step successfully. At each depth, we used the relative ratios of spin coherence times before and after applying external spins on the surface to present the decoherence, and investigated the relationships between depth and ratios. The values of relative ratios declined and then rised with the decreasing depth, which was attributed to the decoherence influenced by external spins, surface spins, discrete surface spin effects and electric field noise. Moreover, our work revealed a characteristic depth at which the NV center would experience relatively the strongest decoherence caused by external spins in consideration of inevitable surface spins. And the characteristic depth was found depending on the adjacent environments of NV centers and the density of surface spins.
\end{abstract}

\pacs{Valid PACS appear here}
\maketitle


\section{\label{sec:level2}Introduction}

The negatively charged nitrogen-vacancy (NV) center in diamond has attracked broad interest owing to its prominent properties. One particular property is the long room-temperature spin coherence time, which is essential for NV centers being used in various applications. Recently, the use of NV centers as sensors to detect external spins \citep{staudacher2013nuclear,mamin2013nanoscale,muller2014nuclear,grotz2011sensing,shi2015single,Panich2011Proton} has been demonstrated, and both the coherence time and detected signal strength are found to be critically dependent on the depth of the NV center \citep{staudacher2013nuclear,mamin2013nanoscale,Grinolds2013Nanoscale,Rugar2014Proton,H2015Nanoscale,Myers2014Probing,Wang2016Coherence}. In view of this, the depth dependent properties of NV centers as well as the preparation methods of NV centers of different depths have been widely investigated.

Conventionally, low-energy nitrogen-implantation\citep{Pezzagna2010Creation} and the epitaxial growth of a high quality nitrogen-doped CVD diamonds followed by electron \citep{Ohno2012Engineering} or ion irradiation \citep{Ohno2014Three} are the methods to make shallow NV centers. Moreover, in order to bring an NV center closer to the diamond surface step by step to investigate the depth dependence of its properties, thermal oxidation\citep{loretz2014nanoscale,Wang2016Coherence,Kim2014Effect} and plasma etching \citep{santori2009vertical,Kim2014Effect,cui2015reduced,favaro2015effect} methods have been developed and widely used in the recent years. Compared with thermal oxidation which performs an etching on the entire diamond sample surface, plasma etching method makes it possible to etch a specific area of the sample by previously depositing a mask on the surface \citep{favaro2015effect}.

In this paper, we performed plasma etching on a bulk diamond to precisely control the depth of NV centers with respect to the sample surface. Then we studied depth dependence of spin coherence times of the NV centers for samples with external nuclear or electronic spin baths around the surface. In particular, by using NV center array and position marks, we could track the very same single NV center at different depths, which enabled us to keep a stable internal adjacent environment of the center and made the depth and external spins the only two variables.

\section{\label{sec:level2}Methods}

We used an electronic grade (100)-oriented, 2 \texttimes{} 2 \texttimes{}  0.5
mm\textsuperscript{3} sized diamond substrate from Element Six ({[}\textsuperscript{13}C{]}{}={}1.1\%,
{[}N{]}{}<{}5{}ppb) for the experiments. By using electron beam lithography,
an arrary made of 60 nm diameter apertures, enclosed with 2 $\mu$m
wide vacant strips (serving also as position marks \citep{wang2015high,Wang2016Coherence}), were patterned on a
300 nm thick polymethyl methacrylate (PMMA) layer previously deposited
on the diamond plate surface \citep{spinicelli2011engineered}. The NV center array in the diamond
was created by ion implantation with the \textsuperscript{14}N\textsubscript{2}\textsuperscript{+}
molecule energy of 50 keV and a fluence
of 0.65 \texttimes{} 10\textsuperscript{11} \textsuperscript{14}N\textsubscript{2}\textsuperscript{+}
 per cm\textsuperscript{2} through the apertures and strips
on the PMMA layer. The implanted sample was annealed at 1050 \textcelsius{}
in vacuum at 2 \texttimes{} 10 \textsuperscript{-5} Pa for 2 h to form
long spin coherence time centers \citep{yamamoto2013extending}. Then the sample, after oxidation
for 2 h in air at 430 \textcelsius , was cleaned with acidic mixture
(sulfuric, nitric, and perchloric acid in a 1 : 1 : 1 ratio) at 200
\textcelsius{} for 1.5 hours.

\begin{figure}[!h]
\centering
  \includegraphics[height=6.5cm]{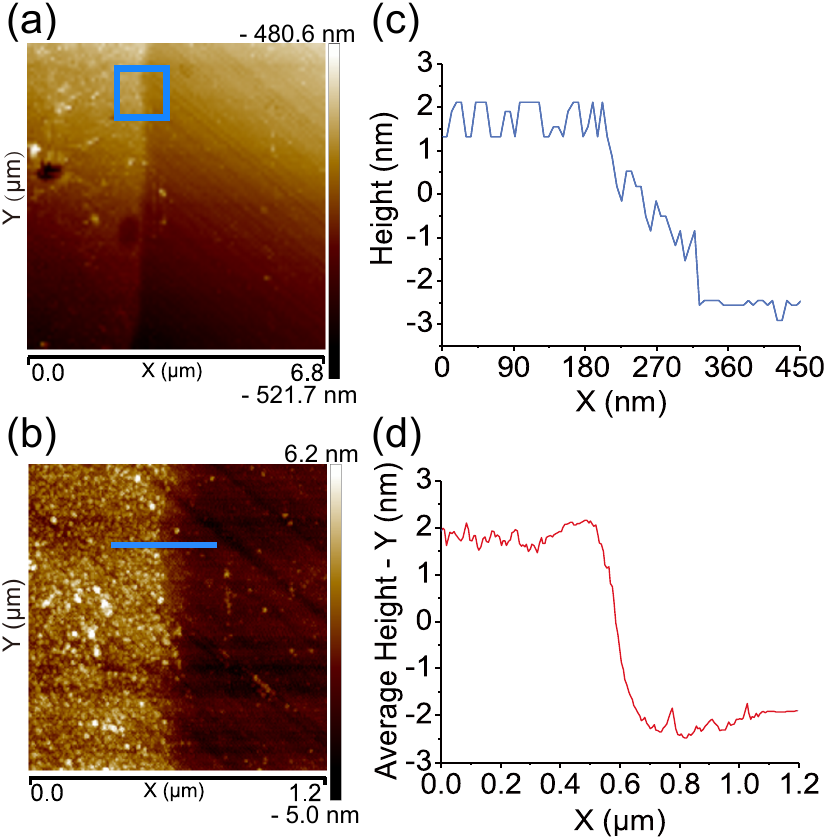}
\caption{Surface morphology of the etched diamond sample characterized by AFM. (a) Image of a representative area of the substrate etched for 20 seconds. (b) Magnified image of the square area delineated in (a). (c) The distribution of height along the blue line in (b). (d) The distribution along X axis of the Y-axis-average height of the whole area in (b).}
\end{figure}
 
The plasma-related processes were performed using an ICP RIE reactor
(Oxford PlasmaPro NGP80 machine equipped with ICP source). In order
to obtain an efficient etching rate, we performed the plasma etching
on the diamond sample in conditions of 200 W ICP power, 30 mTorr chamber
pressure, 10 sccm of oxygen, 5 sccm of argon, which was different
from that used in the previous works \citep{Kim2014Effect,cui2015reduced,favaro2015effect}. By depositing a mask (lithography-patterned AZ 6112 photoresist) on a part of the sample's surface to protect
it from etching, we could get a reference point of the initial depth, with respect to which we could determine the etching rate and depth from surface topography analysis with the atomic force microscope (AFM). Fig.1 demonstrates a representative result for sample of 20 s etching time. Fig.1(a)
shows a 6.8 \texttimes{} 6.8 $\mu$m\textsuperscript{2} area
of the etched diamond sample surface, which demonstrates a clear boundary near the middle left. A plane fit was performed on the 1.2 \texttimes{} 1.2 $\mu$m\textsuperscript{2} square delineated in Fig.1(a) to eliminate the effect of slant angle caused by placing
sample non horizontally.The resulting corrected image is presented in Fig.1(b).
The distribution of height along the randomly selected blue line in
Fig.1(b) is shown in Fig1.(c), which indicates an etching depth
of about 4 nm at the boundary. After removing remarkable spikes and streaks, the distribution along
X axis of the Y-axis-average height is shown in Fig.1(d) from which
an etching depth of about 3.8 nm can be obtained. Deriving the etching
depths for ten etching experiments of different etching times, we
found that the etching rate under the conditions mentioned above was
11.8 \textpm{} 1 nm/min. Therefore, we determined to perform a 20-second
plasma etching on the sample (corresponding to an etching depth of
about 4 nm) each time when the NV centers were distant from the surface, and
a short etching time (corresponding to the etching depths of about 2 nm or 1 nm) when
they were shallow. In this way, we made the centers approaching to
the sample surface step by step until the centers disappeared, from
which the initial depths of NV centers could be derived.

\begin{figure}[!h]
\centering
  \includegraphics[height=6.8cm,width=8.8cm]{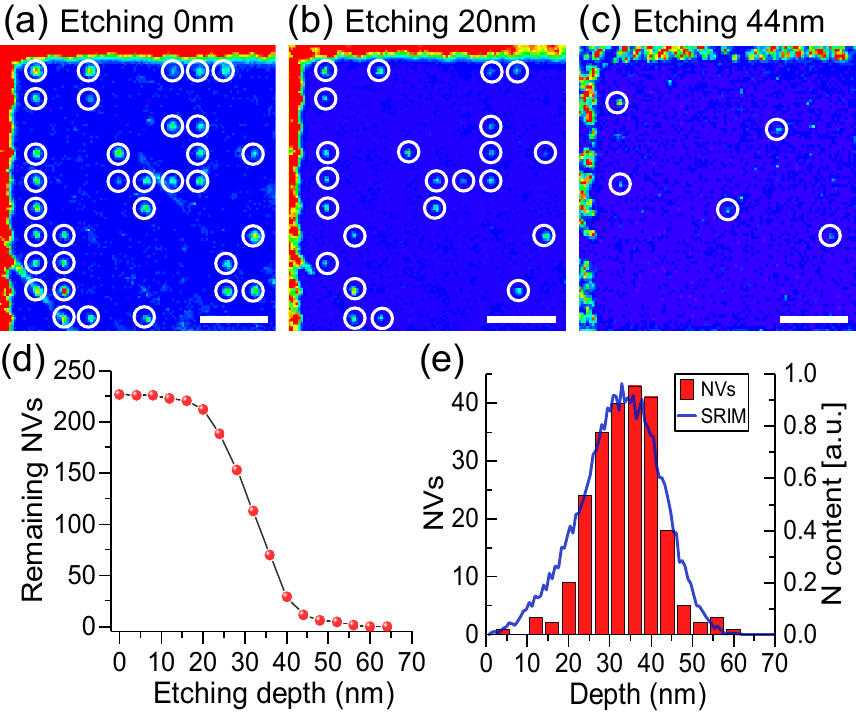}
  \caption{(a), (b) and (c) fluorescence images of the remaining NVs at different etching depths (all with the 5 $\mu$m scale bar). (a) Image of a representative area of the sample before plasma etching. (b) and (c) The same area after plasma etching for 20 nm and 44 nm, respectively. The tracked single centers are circled. (d) The number of remaining NV centers in the tracked area at different etching depths. (e) The distribution histogram of the number of NVs of different depths, and the SRIM simulated depth profiles (blue curve) of the 25 keV implanted nitrogen atoms.}
\end{figure}

We tracked 227 single NV centers to investigate the distribution of center depth. Respectively, Fig.2(a), (b) and (c) show the fluorescence images of the same representative region of the tracked area of the sample etched for three different etching times (corresponding to the etching depths 0 nm, 20 nm, 44 nm). The strip-shaped bright regions on the left and up sides of the images, implying where NV center clusters exist, correspond to the position marks, and the lightspots in white circles represent the tracked single NV centers. Obviously, after etching 20 nm, the position mark strips became less bright, and only 23 centers among the initially tracked 33 single NV centers remained discernible. And after etching 44 nm, the bright strips became interrupted and much less bright, and only 5 single NV centers remained. The evolution of the remaining number of all 227 tracked single NV centers are shown in Fig.2(d). We found that the number of remaining centers reduced slightly when etching depth was 20 nm or less. However, when the etching depth increased further, in a range from 20 to 40 nm, the number reduced dramatically. Finally, when the etching depth was above 40 nm, the number  reduced slowly again. By subtracting the numbers of adjacent etching depths, the distribution histogram was obtained and  shown in Fig.2(e) which corresponded to the SRIM simulation for an implantation nitrogen atom energy of 25 keV. The result supported our estimated etching rate mentioned above.

\section{\label{sec:level3}Results}

\begin{figure*}[htb!]
\centering
  \includegraphics[height=10.4cm,width=16cm]{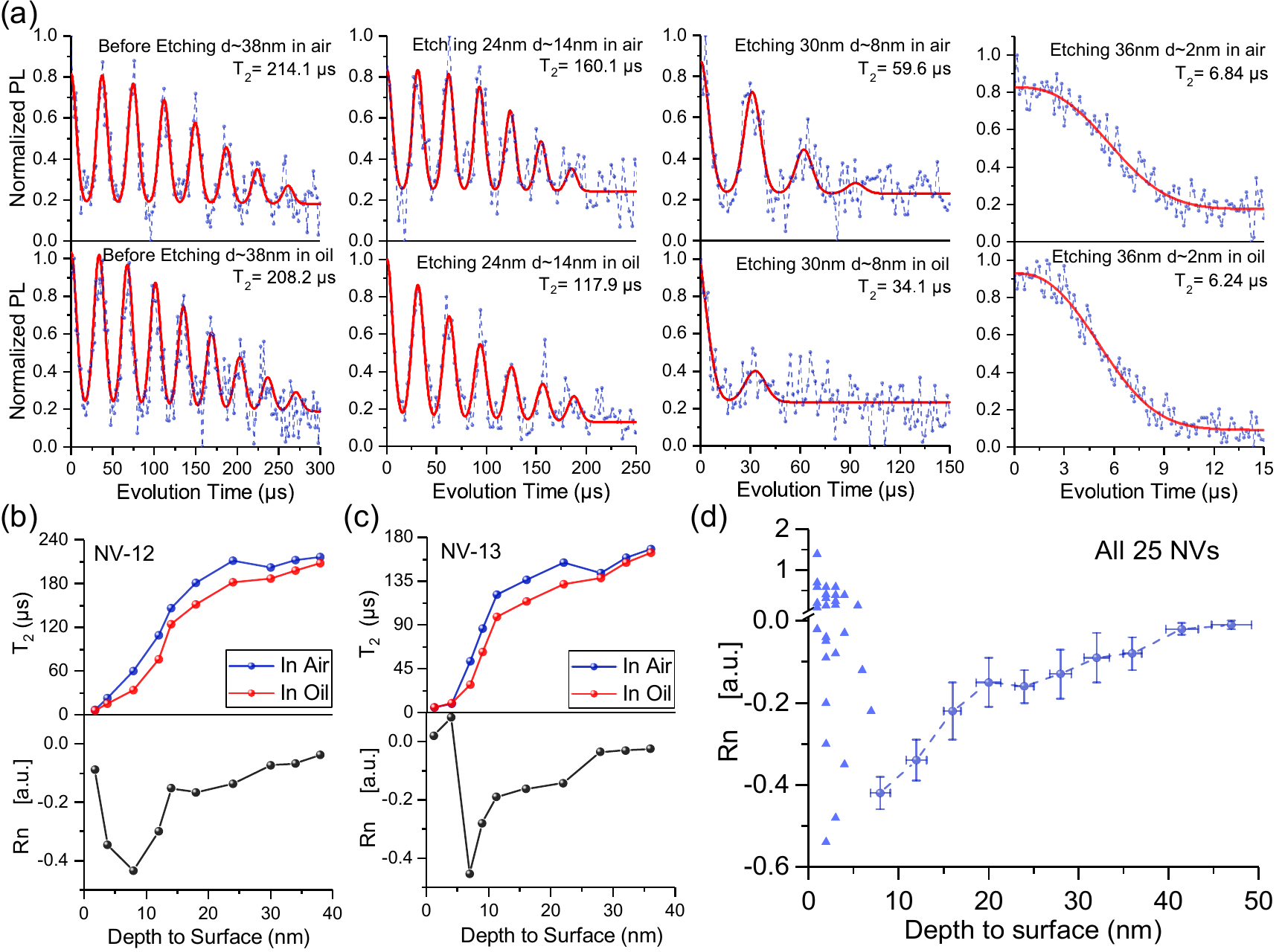}
  \caption{Coherence time measurements of NV centers at different
etching depth before and after applying oil. (a) Spin echo measurements
of NV-12 at four different etching depths, with surface in air atop,
in oil at bottom. (b) and (c) The coherence times of NV-12 and NV-13,
respectively, for various depths to the sample surface before and after oil
applied, atop; and the Rn at bottom. (d) The Rn of all 25 measured NV centers \emph{vs.
}depths to the sample surface (A few data points under 6 nm are overlapped).}
\end{figure*}

At each depth, spin coherence times were measured both before
and after applying external spins to the diamond surface. The two
liquids used were microscope immersion oil and
Cu\textsuperscript{2+} solution, providing external nuclear\citep{pham2016nmr} and electronic\citep{Panich2011Proton}
spins respectively (Though Cu\textsuperscript{2+} provides nuclear spins as well, the electronic spin leads the main effect for decoherence due to the higher gyromagnetic ratio). The sample was placed in custom built confocal microscope system with the applied magnetic field (B = 55 \textpm{} 5
G) paralleling to the detected single NV center axes. Twenty five
single centers (labeled NV-01 \ensuremath{\sim} 25) were randomly
selected for the measurements, and initially, the T\textsubscript{2}
of three centers of them were less than 50 $\mu$s while of
the rest were between 120 to 250 $\mu$s, indicating that most
of the selected centers were deep inside the sample with a spin bath environment of \textsuperscript{13}C impurities \citep{ryan2010robust}.

\begin{figure*}[htbp]
\centering
  \includegraphics[height=10.4cm,width=16cm]{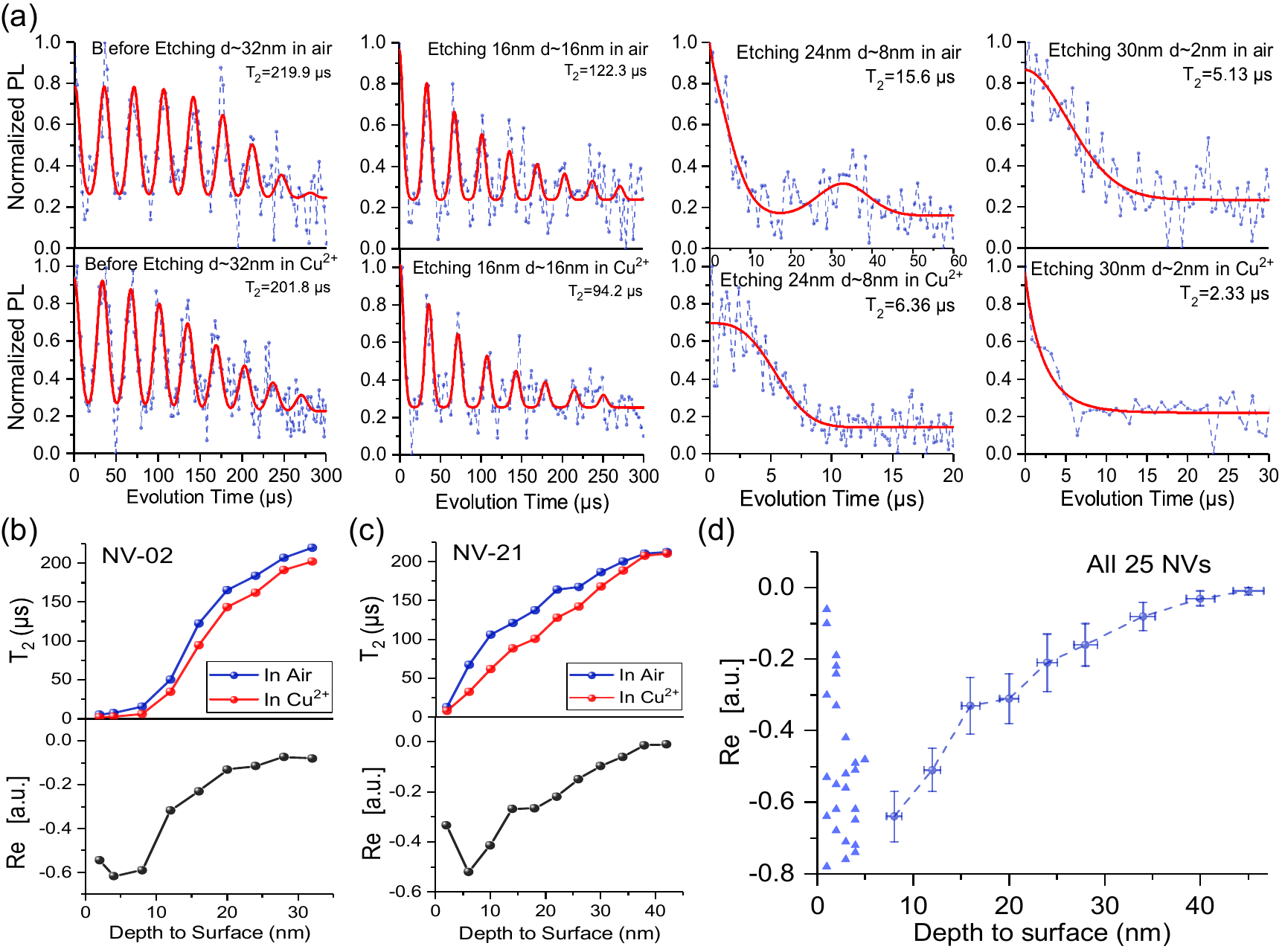}
  \caption{Coherence time measurements of NV centers at different
etching depth before and after applying Cu\textsuperscript{2+} solution.
(a) Spin echo measurements of NV-02 at four different etching depths,
with surface in air atop, in Cu\textsuperscript{2+} solution at bottom.
(b) and (c) The coherence times of NV-02 and NV-21, respectively,
for various depths to the sample surface before and after Cu\textsuperscript{2+}
solution applied, atop; and Re at bottom. (d) All 25 measured NV centers' 
Re \emph{vs. }depths to the sample surface.}
\end{figure*}

Fig.3 exhibits the results of coherence times of the sample with microscope
immersion oil applied on the surface. It has been acknowledged that
when the surface is exposed to air, the coherence time will decline
as NVs approaching to the surface, owing to the influence of surface
spin baths existing naturally \citep{Myers2014Probing,ofori2012spin,rosskopf2014investigation,romach2015spectroscopy} (Since it was difficult to remove the surface spins completely). The coherence times obtained with no external liquid applied
to the surface would be called T\textsubscript{2}\textsuperscript{air} (which served
as a reference), and that with oil applied would be called T\textsubscript{2}\textsuperscript{oil}. Noteworthily, we used the ratio Rn = (T\textsubscript{2}\textsuperscript{oil} - T\textsubscript{2}\textsuperscript{air}) / T\textsubscript{2}\textsuperscript{air} to represent the decoherence caused by external nuclear spins, which reflected the intensity of interaction between NV and external nuclear spins with respect to the inevitable intrinsic spins around the diamond surface. Fig.3(a) shows four representative results of spin echo measurements of NV-12 at different depths. This center disappeared after etching 38 nm, indicating its initial depth to be d \ensuremath{\sim} 38 nm. Before etching, the T\textsubscript{2}\textsuperscript{air} of NV-12 was 214.1 $\mu$s. After applying oil to the surface, the coherence time T\textsubscript{2}\textsuperscript{oil} changed to 208.2 $\mu$s, and Rn was about -0.03, suggesting that the external nuclear spins had
little influence on the coherence time of the center located at a depth about 38 nm. Then, after etching 24 nm (d \ensuremath{\sim} 14 nm), the T\textsubscript{2}\textsuperscript{air} declined to 160.1 $\mu$s,
and T\textsubscript{2}\textsuperscript{oil} declined to 117.9 $\mu$s, making 
Rn to be -0.26, indicating the external nuclear spins indeed had huge
effect on coherence time when NVs became shallow. Even more, T\textsubscript{2}\textsuperscript{air}
and T\textsubscript{2}\textsuperscript{oil} of the same center declined to 59.6 and
34.1 $\mu$s, respectively, for sample etched for 30 nm, which
corresponded to a depth of NV-12 d \ensuremath{\sim} 8 nm. The Rn
reduced to -0.43 which meant the external nuclear spins caused strong
decoherence. Then, another 30-second-etching was performed to make
NV-12 only about 2 nm to the surface, which caused T\textsubscript{2}\textsuperscript{air} 
declined to 6.84 $\mu$s, T\textsubscript{2}\textsuperscript{oil} to 6.24 $\mu$s
However, we found that for the sample thus treated, the Rn increasd
to -0.09. The T\textsubscript{2}\textsuperscript{air} and T\textsubscript{2}\textsuperscript{oil} of
NV-12 for various depths to the sample surface are presented in Fig.3(b)
top, and they both have similar evolution with decreasing depth. As
shown in the figure, both T\textsubscript{2}\textsuperscript{air} and T\textsubscript{2}\textsuperscript{oil}
decreased slowly at the depth above 20 nm, then decreased rapidly
until NV-12 finally disappeared, which was the same as our previous work \citep{Wang2016Coherence}. Besides, the T\textsubscript{2}\textsuperscript{oil}
is less than T\textsubscript{2}\textsuperscript{air}, revealing the external nuclear
spins have enormous influence on coherence time in most cases in addition to intrinsic surface spins. The Rn for the center with various depths are presented in Fig.3(b) bottom, which shows that the Rn curve
goes down with the center depth reducing till about 8 nm and then rises
. Another representative result of coherence time measurements
of single center NV-13 is shown in Fig.3(c). Similar to NV-12, both
T\textsubscript{2}\textsuperscript{air} and T\textsubscript{2}\textsuperscript{oil} of NV-13 decreased
slowly with the decrease of its depth in the range above 15 nm, and
decreased rapidly in the last 20 nm etching. However, the values of Rn at the depths
very near the surface were positive, 0.03 and 0.12 (meaning T\textsubscript{2}
increased when external nuclear spins applied), suggesting the decoherence
was suppressed by applying oil on the surface. 

We summarized the depth dependent Rn of all 25 NV centers in Fig.3(d). For depth above 6 nm,
we divided the depth range into sequential intervals (4 nm for the first 8 intervals, and 6 nm for the last 3 intervals), and figured out the mean Rn in every interval. The obtained mean Rn declined from -0.01 to -0.42 with the depth decreasing. For depth under 6 nm, the data of all 25 NVs scattered between -0.54 and 1.43, most around 0.2, revealing that the decoherence caused by applied oil was unsteady, even suppressed when NVs were very near the surface. The result will be discussed in detail in the following part.

Fig.4 exhibits the results of coherence times with Cu\textsuperscript{2+}
solution (providing external electronic spins) applied on the sample
surface. Likewise, the ratio Re = 
(T\textsubscript{2}\textsuperscript{Cu\textsuperscript{2+}} - T\textsubscript{2}\textsuperscript{air}) / T\textsubscript{2}\textsuperscript{air} was used
to represent the variation of coherence times caused by external electronic spins. Spin echo measurements of NV-02 with four representative etching depths are shown in Fig.4(a). The center NV-02
initially sat at d \textasciitilde{} 32 nm, according to the fact
that it disappeared after etching 32 nm. Before etching, the T\textsubscript{2}\textsuperscript{air} of NV-02 was 219.9 $\mu$s. After applying Cu\textsuperscript{2+} to the sample surface, T\textsubscript{2} ( T\textsubscript{2}\textsuperscript{Cu\textsuperscript{2+}})
changed to 201.8 $\mu$s, and Re was about -0.08. After etching 16 nm (d \ensuremath{\sim}
16 nm), the T\textsubscript{2}\textsuperscript{air} declined to 122.3 $\mu$s,
and T\textsubscript{2}\textsuperscript{Cu\textsuperscript{2+}} declined to 94.2 $\mu$s,
making Re to be -0.23. T\textsubscript{2}\textsuperscript{air} and T\textsubscript{2}\textsuperscript{Cu\textsuperscript{2+}} declined to 15.6 and 6.36 $\mu$s respectively when NV-02 was
etched about 8 nm to the surface. Strong decoherence caused by external
electronic spins was detected as indicated by the reduced Re of -0.59. However,
with etching depth being 30 nm in total, NV-02 was only about 2 nm
to the surface, and T\textsubscript{2}\textsuperscript{air} and T\textsubscript{2}\textsuperscript{Cu\textsuperscript{2+}}
declined to 5.13 and 2.33 $\mu$s, respectively, making Re changed to -0.55. The T\textsubscript{2}\textsuperscript{air} and T\textsubscript{2}\textsuperscript{Cu\textsuperscript{2+}}
of NV-02 for various depths to the sample surface are presented in
Fig.4(b) top, and the ratios (Re) of them are in Fig.4(b) bottom.
Fig.4(c) shows the results of NV-21. Both figures show that
the coherence times declined dramatically when NV centers are under
20 nm, for sample surface with or without Cu\textsuperscript{2+}
applied. Furthermore, T\textsubscript{2}\textsuperscript{Cu\textsuperscript{2+}} was
less than T\textsubscript{2}\textsuperscript{air} at each depth, leading to the fact
that Re was negative, indicating decoherence caused by external
electronic spins was always existing. We also found that Re of
the two NVs increased when they were brought very near the surface by etching the sample surface.
Re of all the 25 single NV centers at different depths were demonstrated
in Fig.4(d).Similar to the processing of Rn, the data of Re in the figure were also divided into two parts, with depths above and under 6 nm. Compared with the evolution of Rn, Re behaved similarly at the depths above 6 nm, also declined regularly. However, when the depth was less than 6 nm, the data points of Re mostly distributed around -0.6, with a few located in the range from -0.42 to -0.06. 

\section{\label{sec:level4}Discussion}

The results demonstrated above showed that the interaction between
NV centers and external spins varied with NV center depth. We took
the evolution of Rn as an instance to elaborate the various reasons
leading to the results. For the NV centers in diamond, it is inevitable
that surface spin bath exists, which contributes to the decoherence
of NV centers along with spins in bulk \citep{Myers2014Probing,ofori2012spin,rosskopf2014investigation,romach2015spectroscopy}. Correspondingly, the coherence time is as follows :

\begin{equation}
T_{2}^{air}=1/[\gamma_{NV}^{2}(\overline{B_{bulk}^{2}}\tau_{bulk}+\overline{B_{surf}^{2}}\tau_{surf})+\frac{1}{2T_{1}}]
\end{equation}

If external spins are applied on the surface of diamond, the decoherence
would also be influenced by the external spins, in addition to the
surface spins and spins in bulk. Thus, T\textsubscript{2} with external spins around the sample surface (T\textsubscript{2}\textsuperscript{ext}) can be written in the
form :

\begin{equation}
T_{2}^{ext}=1/[\gamma_{NV}^{2}(\overline{B_{bulk}^{2}}\tau_{bulk}+\overline{B_{surf}^{2}}\tau_{surf}+\overline{B_{ext}^{2}}\tau_{ext})+\frac{1}{2T_{1}}]
\end{equation}

In the above expressions, $\gamma_{NV}$ is the gyromagnetic ratio
of NV center, and $\overline{B_{bulk}^{2}}$, $\overline{B_{surf}^{2}}$,
$\overline{B_{ext}^{2}}$ are the MS magnetic field signal produced
by internal (bulk), surface and external spins respectively, and $\tau_{bulk}$,
$\tau_{surf}$, $\tau_{ext}$ are the internal (bulk), surface and external
spin baths' autocorrelation times respectively, which can be regarded
as constant parameters at a given temparature. Noteworthily, the contribution
of $\frac{1}{2T_{1}}$ to $\frac{1}{T_{2}}$ is far less than that
of various spin noises, so it can be neglected. Then, the expression
of ratio can be obtained from Eq.(1) and Eq.(2) as :

\begin{equation}
R=\frac{T_{2}^{ext}-T_{2}^{air}}{T_{2}^{air}}=-\frac{\overline{B_{ext}^{2}}\tau_{ext}}{\overline{B_{bulk}^{2}}\tau_{bulk}+\overline{B_{surf}^{2}}\tau_{surf}+\overline{B_{ext}^{2}}\tau_{ext}}
\end{equation}

\noindent It is worth mentioning again that, by using NV center array and position
marks, we could track each paricular NV center no matter how the sample
surface was etched and the external spins were applied. For each tracked
NV center, the internal adjacent environment was unchanged as the
center depth is not very shallow. Therefore, the quantity $\overline{B_{bulk}^{2}}\tau_{bulk}$
in Eq.(3) was constant for the tracked NV center, and would be denoted
by C\textsubscript{bulk} hereafter. Moreover, for each kind of spins,
the autocorrelation time is invariable, i.e. the $\tau_{surf}$ and
$\tau_{ext}$ in Eq.(3) can be regarded as constants (independent
of center depth) as well.

Then we focused on the center depth dependent $\overline{B_{surf}^{2}}$
and $\overline{B_{ext}^{2}}$. When NVs are distant from the surface,
the relationship between $\overline{B_{surf}^{2}}$ and center depth can
be well described based on a model of a 2D layer of surface \emph{g
}= 2 spins in case with the surface exposed to air \citep{Myers2014Probing}. In particular,
for the (100)-oriented diamond, we have :

\begin{equation}
\overline{B_{surf}^{2}}=\sigma(\frac{g\mu_{0}\mu_{B}}{4\pi})^{2}(\frac{{3\pi}}{8d^{4}})
\end{equation}

\noindent where d is the NV center depth and $\sigma$
is the mean surface spin density. We note that besides the variable
d and a unknown quantity of $\sigma$ , the other parameters
in Eq.(4) can be combined as a constant, C\textsubscript{surf}. For
the $\overline{B_{ext}^{2}}$ , using a model of the sample surface
covered with liquid of infinite thickness that provides homogeneous
external nuclear spins \citep{pham2016nmr}, it can be derived: 

\begin{equation}
\overline{B_{ext}^{2}}=\overline{B_{oil}^{2}}=\rho(\frac{\mu_{0}\text{\textcrh}\gamma_{n}}{4\pi})^{2}(\frac{3\pi}{4d^{3}})
\end{equation}

\noindent where $\rho$ is the nuclear spin number density,
d is the NV center depth, and $\gamma_{n}$ is nuclear
gyromagnetic ratio. The unknown quantity in Eq.(5) is $\rho$, the
value of which depends on the applied oil. However, the rest parameters
besides variable d can also be treated as a constant, C\textsubscript{ext}.
With Eq.(4) and (5) substituted into Eq.(3), a simplified expression
of Rn can be obtained :

\begin{equation}
Rn=-\frac{\rho\tau_{ext}C_{ext}/d^{3}}{C_{bulk}+\sigma\tau_{surf}C_{surf}/d^{4}+\rho\tau_{ext}C_{ext}/d^{3}}
\end{equation}

Eq.(6) showed the relationship between Rn and NV center depth. With
the estimated values of C\textsubscript{in}, $\sigma$$\tau_{surf}$
and $\rho$$\tau_{ext}$ in a reasonable
range, Eq.(6) was found to fit the data well. The results of Rn and Re (averiged in each interval) are
demonstrated in Fig.5. It is noticed that the data points of Rn at the depth above 10 nm conform
to the simulated curve in Fig.5(a), while the data points under 10
nm show a deviation from the simulated curve. Moreover, it can be
learnt from Eq.(6) that the positive Rn is nonexistent, which is incompatible with the experiment results when NVs were brought near to the surface. The discrepancy was attributed to the following effects.
Firstly, the surface spin baths can not be simply treated as a 2D
uniform layer when NVs are very shallow (d \ensuremath{\le} 10 nm)
owing to the existence of discrete surface spin effect or spin clustering \citep{Myers2014Probing}, 
which makes the surface spin density inhomogeneous, and consequently,
Eq.(4) is no longer valid. In this case, $\overline{B_{surf}^{2}}$
becomes sensitive to the surface spin distribution and the denominator value of Eq.(6) fluctuates, leading to the Rn value scattering around the simulated curve. Furthermore, recent works
reveal that electric field noise plays a significant role in decoherence of near-surface NV centers \citep{kim2015decoherence,jamonneau2016competition},  
so the fact that Rn can have positive values is related to the
microscope immersion oil, nonconductor of high dielectric constant
($\kappa$ = 2.3), which reduces the electric
field noise and suppresses the decoherence \citep{kim2015decoherence}.

\begin{figure}[!h]
\centering
  \includegraphics[height=4cm,width=8.8cm]{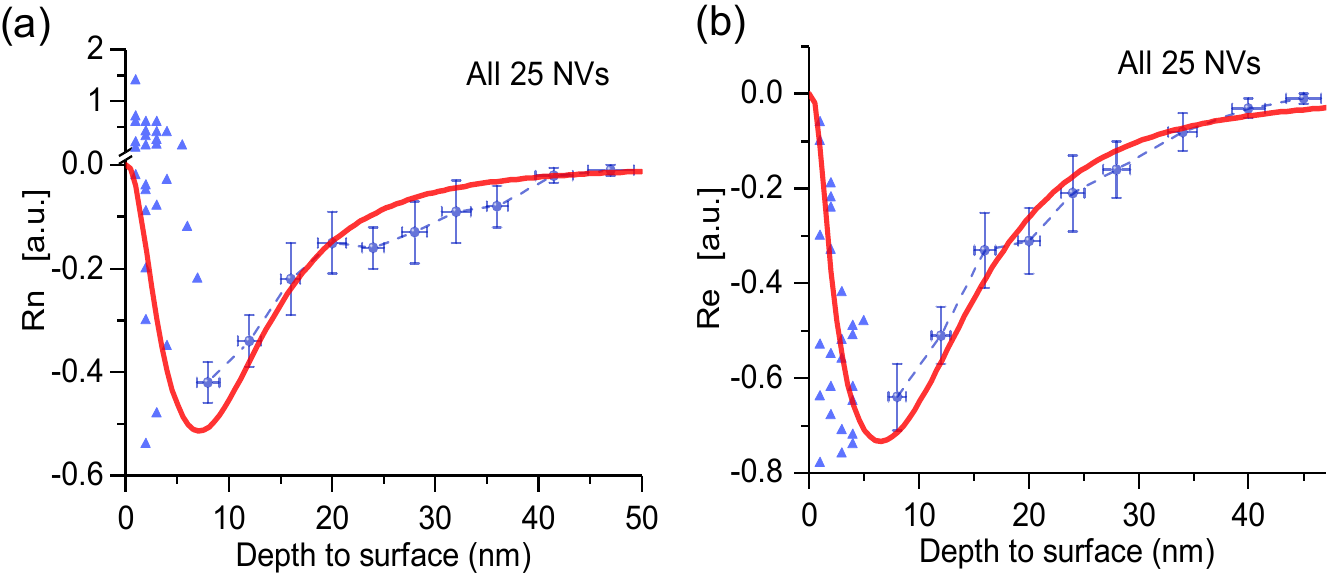}
  \caption{Coherence time ratios of (a) Rn and (b) Re \emph{vs.
}NV center depths to the sample surface. The solid lines in the figure are simulated curves, which are in red online.}
\end{figure}

The expression of the ratio Re (T\textsubscript{2}\textsuperscript{Cu\textsuperscript{2+}} - T\textsubscript{2}\textsuperscript{air}) / T\textsubscript{2}\textsuperscript{air}
is similar to Eq.(6), and the simulated curve presented in Fig.5(b)
demonstrates an appropriate fit with the data. In spite of this,
the data deviation from the simulated curve is shown at the depth
under 5 nm as well, suggesting that the discrete surface spin effect
does influnce the values of the ratios. A difference from Rn is that
no positive Re is found, which can be attributed to that the plentiful ions in Cu\textsuperscript{2+} solution near the sample surface would enlarge the electric field \citep{Newell2016Surface}, instead of decreasing the electric field noise that would suppress the decoherence.

Our results demonstrated that Eq.(6) could explain the depth dependent decoherence behaviours
caused by external spins with respect to the inevitable intrinsic
spins around the diamond surface. Each simulated curve in Fig.5 shows
a minimum at the depth about 6 nm. As an example, by taking the derivative
of Rn in Eq.(6) with respect to the center depth d, it was found that
Rn would take minimum value at the depth d\textsubscript{0} :

\begin{equation}
d_{0}=\sqrt[4]{\frac{\sigma\tau_{surf}C_{surf}}{3C_{bulk}}}
\end{equation}

Eq.(7) reveals that the characteristic depth (d\textsubscript{0}),
at which the ratio Rn (and Re as well) meets the minimum, only depends
on the the adjacent environments of NV centers and the density of
diamond surface spins. In external spin detection, the NV center should
be brought closing to the sample surface as far as possible to strengthen
the detected signal. However, taking surface spins into consideration,
the depth at which the external spins cause relatively the most intense decoherence
is several nanometers away from the surface. Since the adjacent environment
of NV center is changeless, it can be obtained from Eq.(7) that d\textsubscript{0 }decreases
with $\sigma$ reducing. Therefore, decreasing the density of
surface spins by proper surface treatments can lower d\textsubscript{0}
and hence the external spin detection with less influence of surface spins
can be realized for the shallow NV center sensors.

\section{\label{sec:level5}Summary}

We investigated the depth dependence of the coherence
times of NV centers for diamond plate with or without external nuclear
and electronic spins around the surface. By using NV center array and
position marks, each particular NV in the diamond plate etched for
different depths could be recognized and tracked. As the internal
adjacent environments of the tracked NVs were kept unchanged upon etching, our results obtained by 
T\textsubscript{2} tracking was more persuasive than that by measuring NVs
initially in different depths. We performed plasma etching to control
the depths of NVs with an efficient etching rate. Based on this, we
applied microscope immersion oil and Cu\textsuperscript{2+} solution
on the surface to obtain external nuclear and electronic spins. We introduced the coherence time ratios of Rn and Re to present the decoherence caused by external spins,
and found the depth dependent ratios behaved in the form of a function having a minimum. The characteristic depth at which the NV centers experienced relatively the strongest decoherence caused by external spins, as indicated by the minimum ratio R = (T\textsubscript{2}\textsuperscript{ext} - T\textsubscript{2}\textsuperscript{air})
 / T\textsubscript{2}\textsuperscript{air}, was found depending on the adjacent environment of NV center and the density of surface spins, which could be useful in the
further study and detection of external spins, surface spin noise and so forth with NV centers in diamond.

\section*{Acknowledgements}

We thank X.X.Wang, J.L.Peng, D.F.Zhou and W.Liu from the USTC Center
for Micro- and Nanoscale Research and Fabrication for the technical
support of Plasma etching and AFM. We also thank G.P.Guo and J.You
from the Key lab of Quantum Information for the support of electron
beam lithography. This work was supported by the National Basic Research Program of China (2011CB921400, 2013CB921800) and the National Natural Science Foundation of China (Grant Nos. 11374280, 50772110).

\bibliographystyle{unsrtnat}

\providecommand*{\mcitethebibliography}{\thebibliography}
\csname @ifundefined\endcsname{endmcitethebibliography}
{\let\endmcitethebibliography\endthebibliography}{}

\end{document}